\documentclass[conference,10pt]{IEEEtran}
\usepackage{epsfig,rotating,setspace,latexsym,amsmath,epsf,amssymb,amsfonts,bm,theorem,subfigure,epstopdf}
\usepackage{cite,authblk}
\usepackage{bbm}
\usepackage[ruled,vlined]{algorithm2e}

\usepackage{color}

\DeclareMathOperator*{\argminA}{arg\,min}

\IEEEoverridecommandlockouts
\allowdisplaybreaks

\begin{document}
\bstctlcite{IEEEexample:BSTcontrol} 


\title{How to Make 5G Communications ``Invisible": \\ Adversarial Machine Learning for Wireless Privacy}
	
\author[1]{Brian Kim}
\author[2]{Yalin E. Sagduyu}
\author[2]{Kemal Davaslioglu}
\author[2]{Tugba Erpek}
\author[1]{Sennur Ulukus}
	
\affil[1]{\normalsize Department of Electrical and Computer Engineering, University of Maryland, College Park, MD 20742, USA}
\affil[2]{\normalsize Intelligent Automation, Inc., Rockville, MD 20855, USA \thanks{This effort is supported by the U.S. Army Research Office under contract W911NF-17-C-0090. The content of the information does not necessarily reflect the position or the policy of the U.S. Government, and no official endorsement should be inferred.}}
	
\maketitle

\begin{abstract}
We consider the problem of hiding wireless communications from an eavesdropper that employs a deep learning (DL) classifier to detect whether any transmission of interest is present or not. There exists one transmitter that transmits to its receiver in the presence of an eavesdropper, while a cooperative jammer (CJ) transmits carefully crafted adversarial perturbations over the air to fool the eavesdropper into classifying the received superposition of signals as noise. The CJ puts an upper bound on the strength of perturbation signal to limit its impact on the bit error rate (BER) at the receiver. We show that this adversarial perturbation causes the eavesdropper to misclassify the received signals as noise with high probability while increasing the BER only slightly. On the other hand, the CJ cannot fool the eavesdropper by simply transmitting Gaussian noise as in conventional jamming and instead needs to craft perturbation signals built by adversarial machine learning to enable covert communications. Our results show that signals with different modulation types and eventually 5G communications can be effectively hidden from an eavesdropper even if it is equipped with a DL classifier to detect transmissions.  

\end{abstract}

\section{Introduction}\label{sec:Introduction}
Information privacy is a fundamental problem in wireless communications due to the open and shared nature of wireless medium. A canonical setting consists of a transmitter-receiver pair and an eavesdropper that aims to infer about the communications from the transmitter to its receiver. The eavesdropper may pursue different objectives such as decoding transmissions or detecting whether there is an ongoing transmission, or not. Information privacy regarding unauthorized decoding has been extensively studied both from encryption-based security and information-theoretical studies \cite{yener-ulukus,Schaefer}. In this paper, we consider an eavesdropper that pursues the second objective, namely detecting an ongoing transmission. Once a transmission is detected, other attacks such as jamming can be launched subsequently. 

\emph{Covert communication} has been  studied in terms of hiding information in noise where the main goal has been to reduce the signal-to-noise ratio (SNR) at the eavesdropper \cite{Bloch, Wang}. A fundamental bound has been established on the total transmit power over a given number of channel uses while keeping communications covert, known generally as the square-root law \cite{Goeckel}; see also \cite{mukherjee-ulukus} for related work. In this paper, we approach the covert communications problem from an \emph{adversarial machine learning} (AML) point of view. AML studies the problem of machine learning (ML) in the presence of adversaries that may attempt to manipulate test and training processes of ML algorithms \cite{Kurakin1, Shi2017, Vorobeychik1}. While the origin of applications on AML is in computer vision, there has been a growing interest in applying AML to wireless communications,  including exploratory (inference), causative (poisoning), evasion, Trojan and spoofing attacks \cite{Shi2018, terpek, Davaslioglu1, Shi, erpek1, Luo2019, Sagduyu2020}.

We consider an eavesdropper with a deep learning (DL) based classifier to identify an ongoing transmission. This classifier achieves high accuracy in distinguishing received signals from noise. 
We introduce a \emph{cooperative jammer (CJ)} that has been extensively used in physical layer security literature \cite{Tekin, xie-ulukus, bassily}. In this paper, the CJ transmits over the air at the same time as the transmitter with the purpose of misleading the eavesdropper's classifier.
This corresponds to an \emph{evasion attack} (or \emph{adversarial attack}) in AML. Evasion attacks have been used to manipulate spectrum access \cite{YiMilcom2018, Yalin2019, Sagduyu1}, autoencoder communications \cite{Larsson1} and modulation classifiers \cite{Larsson2, Kim, Kim2}. In this paper, evasion attack is used as means of covert communications to prevent an eavesdropper from distinguishing an ongoing transmission from noise.


We use the CJ as the source of adversarial perturbation to fool the classifier at an eavesdropper into making classification errors. While a large perturbation added by the CJ can easily fool the classifier, it would also increase the interference and the bit error rate (BER) at the intended receiver to undesirable levels. Therefore, an upper bound on the perturbation strength is imposed. A special case of our setting has been considered in \cite{Hameed}, where the transmitter adds perturbations to its own signals to fool a modulation classifier while aiming to sustain its own communications performance. In this paper, our focus is on covert communications assisted by a CJ, whose position can further boost the impact on the eavesdropper to classify received signal as noise while reducing the impact on the BER performance. Note that it is typically more challenging to fool a classifier into misclassifying a signal as noise (as considered in this paper) instead of confusing it among signal types (as considered in \cite{Hameed}). For the performance evaluation, we start with basic (QPSK and 16QAM) modulated signals, and then extend to more complicated 5G communication signals, and show that we can effectively hide these signals from an eavesdropper that uses a DL classifier to detect transmissions.

The perturbation of the CJ is selected to minimize the strength of the perturbation subject to the condition of successfully fooling the eavesdropper, and an upper bound on the perturbation power that can translate to limiting the BER at the receiver. We show that Gaussian noise is not effective as an adversarial perturbation and develop an algorithm to optimize perturbations for the CJ to enable covert communications that we demonstrate for signals with different modulation types and 5G communications. 

\begin{figure}[t]
	\centerline{\includegraphics[width=0.8\linewidth]{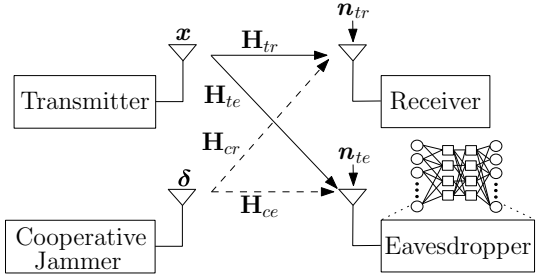}}
	\caption{System model.}
	\label{sys}
\end{figure}

\section{System Model} \label{sec:SystemModel}
We consider a wireless system that consists of a transmitter, a receiver, a CJ, and an eavesdropper as shown in Fig.~\ref{sys}. 
The transmitter sends $p$ complex symbols, $\boldsymbol{x}\in \mathbb{C}^{p}$, by mapping a binary input sequence $\boldsymbol{m}\in \{0,1\}^{l}$. Specifically, $\boldsymbol{x} = g_{s}(\boldsymbol{m})$, where  $g_{s} : \{0,1\}^{l} \rightarrow \mathbb{C}^{p}$ and $s$ represents the modulation type of the transmitter.
Then the transmitter's signal received at node $j$ (either the receiver $r$ or the eavesdropper $e$) is 
\begin{align}
	\boldsymbol{r}_{tj} = \mathbf{H}_{tj} g_{s}(\boldsymbol{m})+\boldsymbol{n}_{tj} = \mathbf{H}_{tj}\boldsymbol{x} + \boldsymbol{n}_{tj}, \quad j \in \{r,e\},
\end{align}
where $\mathbf{H}_{tj} = \mbox{diag} \{h_{tj,1},\cdots, h_{tj,p}\}\in \mathbb{C}^{p\times p}$ and $\boldsymbol{n}_{tj}\in \mathbb{C}^{p}$ are the channel and complex Gaussian noise from the transmitter to node $j$, respectively. Upon receiving the signal $\boldsymbol{r}_{tr}$, the receiver decodes the message with the BER given by
\begin{equation}
\mathrm{P}_{e}(\boldsymbol{m},\boldsymbol{r}_{tr}) = \frac{1}{l}\sum_{i=1}^{l} \mathbb{I}\{m_i \ne \hat{m}_i\}, 
\end{equation}
where $\hat{m}_i$ is a decoded bit and $\mathbb{I}\{\cdot\}$ is an indicator function.

The eavesdropper tries to detect the existence of wireless transmission using a pre-trained DL-based classifier, namely a deep neural network (DNN), $f(.;\boldsymbol{\theta}): \mathcal{X} \rightarrow \mathbb{R}^{2}$, where $\boldsymbol{\theta}$ is the set of DNN parameters and $\mathcal{X} \subset \mathbb{C}^{p}$. Input $\boldsymbol{x}\in \mathcal{X}$ is assigned label $\hat{l}(\boldsymbol{x},\boldsymbol{\theta}) = \arg \max_{k} f_{k}(\boldsymbol{x},\boldsymbol{\theta})$ where $f_{k}(\boldsymbol{x},\boldsymbol{\theta})$ is the output of classifier $f$ corresponding to the $k$th class.

To make communications between the transmitter and its receiver covert, the CJ transmits perturbation signal $\boldsymbol{\delta}\in \mathbb{C}^{p}$ to cause misclassification at the eavesdropper by changing the label of the received signal $\boldsymbol{r}_{te}$ from \emph{signal} to \emph{noise}. Thus, if the transmitter transmits $\boldsymbol{x}$, the received signal at node $j$ is
\begin{align}
	&\boldsymbol{r}'_{tj}(\boldsymbol{\delta}) = \mathbf{H}_{tj}\boldsymbol{x}+\mathbf{H}_{cj}\boldsymbol{\delta}+\boldsymbol{n}_{tj}, \quad j \in \{r,e\},
\end{align}
where $\mathbf{H}_{cj}=\mbox{diag} \{h_{cj,1},\cdot\cdot\cdot, h_{cj,p}\}\in \mathbb{C}^{p\times p}$ is the channel from the CJ to node $j$.

Since the perturbation signal from the CJ not only creates interference at the eavesdropper, but also at the receiver, the CJ determines its signal $\boldsymbol{\delta}$ to cause misclassification at the eavesdropper using fixed power budget $P_{max}$ that also limits the BER at the receiver. Formally, the CJ first determines $\boldsymbol{\delta}$ by solving the following optimization problem:
\begin{align} \label{eq:perturbation}
\argminA_{\boldsymbol{\delta}}& \quad ||\boldsymbol{\delta}||_2\nonumber\\
 s.t. &\quad\hat{l}(\boldsymbol{r}_{te},\boldsymbol{\theta}) \ne \hat{l}(\boldsymbol{r}'_{te}(\boldsymbol{\delta}),\boldsymbol{\theta}) \nonumber \\ &\quad  ||\boldsymbol{\delta}||^{2}_{2} \le P_{max}. 
\end{align}
The solution $\boldsymbol{\delta}^*$ to (\ref{eq:perturbation}) results in the BER,   $\mathrm{P}_{e}(\boldsymbol{m},\boldsymbol{r}'_{tr}(\boldsymbol{\delta}^*))$, at the receiver that can be bounded to a target level by selecting $P_{max}$ accordingly. Since solving (\ref{eq:perturbation}) is difficult, different methods have been proposed in computer vision to approximate the adversarial perturbations like the fast gradient method (FGM) \cite{Kurakin1}. The FGM is computationally efficient for crafting adversarial attacks by linearizing the loss function, $L(\boldsymbol{\theta},\boldsymbol{x},\boldsymbol{y})$, of the DNN classifier in a neighborhood of $\boldsymbol{x}$ where $\boldsymbol{y}$ is the label vector. This linearized function is used for optimization. 
 In this paper, we consider a targeted attack, where the perturbation of the CJ aims to decrease the loss function of the class \emph{noise} and cause a specific misclassification, from \emph{signal} to \emph{noise}, at the eavesdropper even though there is an actual transmission. 
We approach the problem from an AML point of view and aim to fool a target classifier, which is equivalent to hiding communications in noise from a wireless communication perspective. While designing the perturbation, we constrain the BER at the receiver to stay below certain level while satisfying the power constraint at the CJ, as stated in the constraints of the optimization problem (\ref{eq:perturbation}). We assume that the CJ collaborates with the transmitter and thus knows the transmitted signal from the transmitter. 


\section{Adversarial Perturbation for the CJ} \label{sec:jammer}
In this section, we design the white-box perturbation for the CJ using targeted FGM to solve (\ref{eq:perturbation}). For the targeted attack, the CJ minimizes $L(\boldsymbol{\theta},\boldsymbol{r}'_{te}(\boldsymbol{\delta}),\boldsymbol{y}^{target})$ with respect to $\boldsymbol{\delta}$ where $\boldsymbol{y}^{target}$ is one-hot encoded desired target class. We fix $\boldsymbol{y}^{target}$ as \emph{noise} class since the CJ always tries to add perturbation to fool the eavesdropper into misclassifying a received signal as noise. We use FGM to linearize the loss function as $L(\boldsymbol{\theta},\boldsymbol{r}'_{te}(\boldsymbol{\delta}),\boldsymbol{y}^{target}) \approx  L(\boldsymbol{\theta},\boldsymbol{r}_{te},\boldsymbol{y}^{target}) + \boldsymbol{\delta}^{T} \nabla_{\boldsymbol{x}}L(\boldsymbol{\theta},\boldsymbol{r}_{te},\boldsymbol{y}^{target}) $ and then minimize it by setting $\boldsymbol{\delta} = -\alpha \nabla_{\boldsymbol{x}}L(\boldsymbol{\theta},\boldsymbol{r}_{te},\boldsymbol{y}^{target})$, where $\alpha$ is a scaling factor to constrain the adversarial perturbation power to $P_{max}$. After we obtain $\boldsymbol{\delta}$ that causes misclassification at the eavesdropper and satisfies the power constraint, we check the BER at the receiver. The perturbation power can further be adjusted to meet a target BER level.
The details of determining the CJ's perturbation signal are presented in Algorithm \ref{alg1}. 
\setlength{\voffset}{.015in}

\begin{algorithm}[t]
\DontPrintSemicolon
\SetAlgoLined
\label{alg1}
 Inputs: input $\boldsymbol{r}_{te}$, desired accuracy $\varepsilon_{acc}$, power constraint $P_{\textit{max}}$ and $L(\boldsymbol{\theta},\cdot,\cdot)$\\
 Initialize: $ {\varepsilon}\leftarrow {0}, \varepsilon_{\textit{max}} \leftarrow \sqrt{P_{\textit{max}}}, \varepsilon_{min} \leftarrow 0$ \\
 $\boldsymbol{\delta}_{norm} =\frac{\nabla_{\boldsymbol{x}}L(\boldsymbol{\theta},\boldsymbol{r}_{te},\boldsymbol{y}^{\textit{target}})}{(||\nabla_{\boldsymbol{x}}L(\boldsymbol{\theta},\boldsymbol{r}_{te},\boldsymbol{y}^{\textit{target}})||_{2})}$\\
 \While{$\varepsilon_{\textit{max}}-\varepsilon_{min} > \varepsilon_{acc}$}{
  $\varepsilon_{avg} \leftarrow (\varepsilon_{\textit{max}}+\varepsilon_{min})/2$\\
  $\boldsymbol{x}_{adv} \leftarrow \boldsymbol{r}_{te} - \varepsilon_{avg}\boldsymbol{\delta}_{\textit{norm}}$\\
  \lIf{$\hat{l}(\boldsymbol{x}_{adv})== noise$}{
   $\varepsilon_{min}\leftarrow \varepsilon_{avg}$}
  \lElse{$\varepsilon_{\textit{max}}\leftarrow \varepsilon_{avg}$}
   }{
   $\varepsilon = \varepsilon_{\textit{max}}$, $\boldsymbol{\delta}^{jam} = -\varepsilon\boldsymbol{\delta}_{norm} $
  }\\
\caption{Generating the perturbation of the CJ}
\end{algorithm}

%
%

\section{Simulation Setting} \label{sec:SimResults}
We analyze the success of covertness achieved by CJ's perturbation at the eavesdropper and the corresponding effect on the BER at the receiver. We compare this perturbation with random Gaussian noise transmitted by the CJ. Furthermore, we change the location of the CJ to investigate the effects of topology and channel. We assume that the binary source data is generated independently and uniformly at the receiver. The classifier at the eavesdropper is a convolutional neural network (CNN). The input to the CNN is of two dimensions (2,16) corresponding to 16 in-phase/quadrature (I/Q) data samples. The CNN consists of a convolutional layer with kernel size $(1,3)$, a hidden layer with dropout rate $0.1$, ReLu activation function at convolutional and hidden layers and softmax activation function at the output layer that provides the label \emph{signal} or \emph{noise}. We apply backpropagation algorithm with Adam optimizer to train the CNN using cross-entropy as the loss function. The CNN is implemented in Keras with TensorFlow backend. We assume that the eavesdropper already knows the signal type that is used at the transmitter. Thus, the classifier at the eavesdropper is only trained with two labels, \emph{signal} and \emph{noise}. For each signal type, we train a separate classifier using different datasets, where 20000 symbols are generated and split into blocks of 16 I/Q symbols. The channel between the nodes have path-loss effects such that the channel gain from node $i$ to node $j$ is ${h}_{ij}= \left(\frac{d_{0}}{d_{ij}}\right)^{\gamma}$, where $d_{ij}$ is the distance from node $i$ to $j$, $d_0$ is the reference distance and $\gamma$ is the path loss exponent. We set $d_0 = 1$ and $\gamma = 2.8$ throughout the simulations. Note that we can further use methods introduced in \cite{Kim} to incorporate other channel effects such as shadowing and small-scale fading.

We use the perturbation-to-noise ratio (PNR) metric from \cite{Larsson2} that captures the relative perturbation power at the CJ with respect to the noise and measure how the increase in the PNR affects the accuracy of the classifier at the eavesdropper. As the PNR increases, the perturbation generated at the CJ is more likely to be detected by the eavesdropper and increases the BER at the receiver.


\section{Performance Evaluation for Signals with Different Modulations}\label{sec:different modulation}

\subsection{Covertness of Communications}
\begin{figure}[t]
	\centerline{\includegraphics[width=1.0\linewidth]{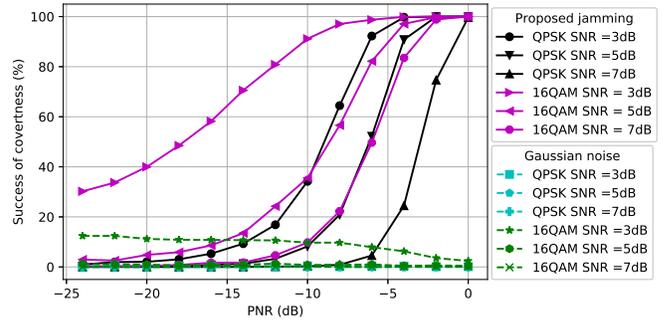}}
	\caption{Success of covertness at the eavesdropper when $d_{ce} = d_{cr}=1$.}
	\label{plot1}
\end{figure}

We aim to hide signals with a fixed modulation scheme, namely QPSK or 16QAM, used by the transmitter. The first topology that we consider is $d_{cr} = d_{ce} = 1$. In Fig. \ref{plot1}, we show how the perturbation signal generated at the CJ affects the classifier at the eavesdropper. The $x$-axis is the PNR (measured in dB) and the $y$-axis is the success of covertness (measured in percentage) that indicates the success of making wireless communication covert, namely the likelihood that the eavesdropper classifies signal plus perturbation as noise. We observe that as the SNR of the signal increases, the CJ needs more perturbation power to cause misclassification at the eavesdropper. Furthermore, the 16QAM-modulated signal is  more susceptible to adversarial perturbation than the QPSK-modulated signal, since it is more difficult to distinguish the 16QAM-modulated signal from the noise for the same SNR. Also, we observe that the success of covertness suddenly increases after some PNR value for both modulation types. On the contrary, the  Gaussian noise-based perturbation has negligible effect on the classifier for all SNR values. We further observe that Gaussian noise with more power decreases the success of covertness when the SNR of 16-QAM modulated signal is 3 dB. The reason is the randomness of the Gaussian noise, which in fact strengthens the noise-like signal to be classified as a signal at the eavesdropper.

\begin{figure}[t]
	\centerline{\includegraphics[width=1.0\linewidth]{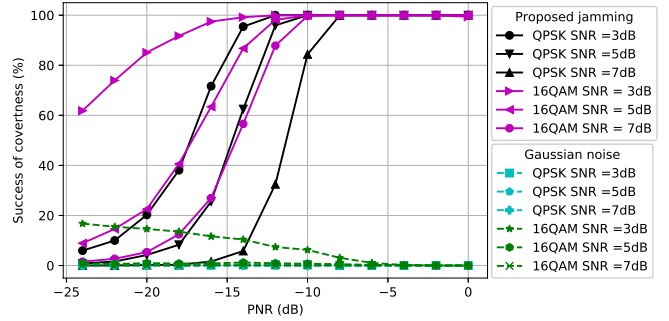}}
	\caption{Success of covertness at the eavesdropper when $d_{ce} = 0.5$ and $d_{cr}=1.5$.}
	\label{plot3}
\end{figure}

In Fig. \ref{plot3}, we consider $d_{cr} = 1.5$ and $d_{ce} = 0.5$ (namely, the distance between the CJ and the receiver is increased and the distance between the CJ and the eavesdropper is decreased compared to Fig. \ref{plot1}). As the SNR of the signal increases, the CJ requires more power to cause misclassification at the eavesdropper, as we have also observed in Fig. \ref{plot1}. Due to the reduced path loss effect between the CJ and the eavesdropper, less power is required to cause misclassification compared to Fig. \ref{plot1}. This result motivates the use of AML instead of conventional jamming (e.g., \cite{Sagduyu2008}) to attack an eavesdropper. 


\subsection{Reliability of Communications}

\begin{figure}[t]
	\centerline{\includegraphics[width=1.0\linewidth]{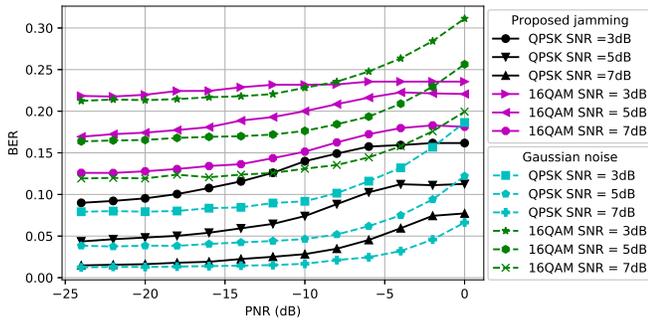}}
	\caption{BER at the receiver when $d_{ce} = d_{cr}=1$.}
	\label{plot2}
\end{figure}

\begin{figure}[t]
	\centerline{\includegraphics[width=1.0\linewidth]{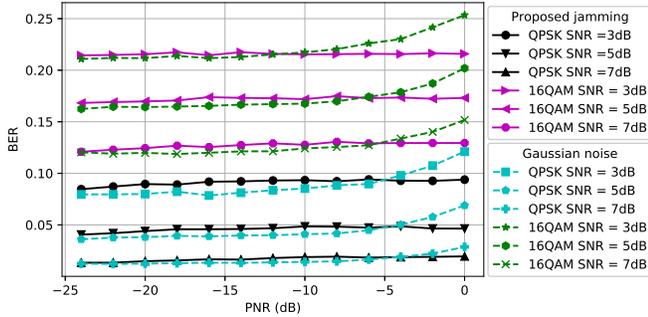}}
	\caption{BER at the receiver when $d_{ce} = 0.5$ and $d_{cr}=1.5$.}
	\label{plot4}
\end{figure}

The BER performance at the receiver for different modulation types and SNR values is compared in Fig. \ref{plot2} when $d_{cr} = d_{ce} = 1$. We observe that the BER of 16QAM-modulated signals is more susceptible to the adversarial perturbation signal than the BER of QPSK-modulated signals. The reason is that since the 16QAM transmits more bits than the QPSK, the distances between constellation points are smaller, which leads to a larger BER for a given SNR. Moreover, as the SNR increases, the average BER decreases as expected. For the CJ with proposed adversarial perturbation, we observe that the BER curve saturates after some PNR value because the successful perturbation signal can be generated using less power than the maximum power that the CJ can use. 
Fig. \ref{plot2} can be used as a guideline to determine the maximum PNR to satisfy the BER requirement at the receiver. For example, to meet the target BER of $0.15$ for a QPSK-modulated signal, the PNR is selected at most -8dB when SNR is 3dB and the resulting success of covertness is 65\%. Furthermore, we observe that the Gaussian noise-based perturbation results in lower BER than the adversarial perturbation in the low PNR regime. However, the BER gap between these two CJ schemes decreases when the PNR increases, and the adversarial perturbation results in a smaller BER in the high PNR region. 


The BER performance at the receiver for different modulation types and SNR values is compared in Fig. \ref{plot4} when $d_{cr} = 1.5$ and $d_{ce} = 0.5$. We observe that the BER gap between the Gaussian noise and adversarial perturbation for the same SNR value decreases due to the increased path loss effect between the CJ and the receiver. Thus, the CJ can create a perturbation signal that causes misclassification with higher success without increasing the BER further if the location of the CJ is closer to the eavesdropper. This result motivates the control of CJ positions to fool a target classifier while protecting the BER performance of the intended receiver.

\section{Performance Evaluation for 5G Communications}
As a full-fledged waveform to hide, we consider the 5G  physical layer communications where a 5G User Equipment (UE) transmits the 5G uplink signal to the base station (gNodeB) in the presence of the perturbation from the CJ. MATLAB 5G toolbox is used to generate 5G signals that include the transport (uplink shared channel, UL-SCH) and physical channel. 
The transport block is segmented after the cyclic redundancy check (CRC) addition and low-density parity-check (LDPC) coding is used as forward error correction. The output codewords are QPSK modulated as an example. Next, the generated resource grid is OFDM modulated with Inverse Fast Fourier Transform and Cyclic Prefix (CP) addition operations where the subcarrier spacing is 15 kHz. 
Target code rate is set to $\frac{820}{1024}$ and the output I/Q samples are stored after the signal passes through the 
channel. Eavesdropper attempts to distinguish the received signals from noise, whereas the receiver attempts to decode the received signals by removing the CP and performing FFT, channel equalization, QPSK demodulation, LDPC and CRC decoding operations. 



\begin{figure}[t]
	\centerline{\includegraphics[width=1.0\linewidth]{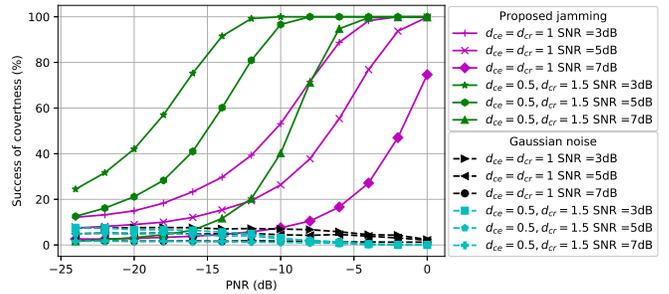}}
	\caption{5G communications covertness performance at the eavesdropper.}
	\label{plot5}
\end{figure}

\subsection{Covertness of Communications}
 The success of covertness for 5G communications is considered in Fig. \ref{plot5}. As we have seen in the previous figures for QPSK-modulated signals and 16QAM-modulated signals, the proposed jamming outperforms the Gaussian noise significantly in high PNR region for 5G communication signals. Also, we observe that more power is needed for CJ to fool the classifier at the eavesdropper when the distance between the CJ and the eavesdropper increases.

\begin{figure}[t]
	\centerline{\includegraphics[width=1.0\linewidth]{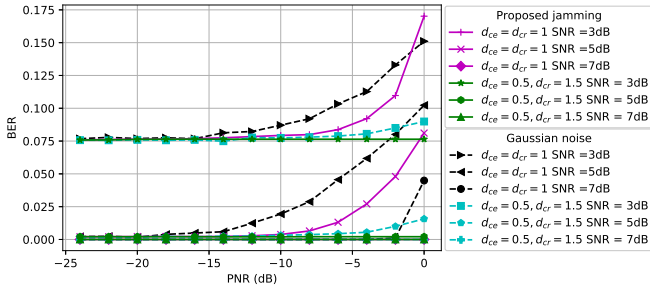}}
	\caption{5G communications BER performance at the receiver. }
	\label{plot6}
\end{figure}

\subsection{Reliability of Communications}
The BER for 5G communications is shown in Fig. \ref{plot6}. When $d_{ce} = d_{cr}=1$ and SNR is 5 dB, the Gaussian noise-based perturbation has higher BER performance compared to the proposed perturbation and a similar result is also observed for other SNR values. Note that the adversarial perturbation by the CJ not only increases the success of covertness, but also has less effect on the BER performance of the receiver compared to the Gaussian noise-based perturbation for 5G communication signals. We further observe that the Gaussian noise-based perturbation results in higher BER than the proposed adversarial perturbation when $d_{ce} = 0.5, d_{cr}=1.5$.

\section{Conclusion} \label{sec:Conclusion}
We considered a wireless communication system in which a CJ designs its perturbation signal to fool a DL-based classifier at the eavesdropper into classifying the ongoing transmissions as noise. Following the AML approach, the CJ was designed to generate the perturbation signal with the FGM method subject to channel effects. For both basic modulated signals and sophisticated 5G signals, we showed that the CJ can generate a perturbation signal that causes misclassification at the eavesdropper (from \emph{signal} to \emph{noise}) with high success, while the BER at the receiver is only slightly affected.
On the other hand, CJ with Gaussian noise is not successful in fooling the classifier. These results demonstrate the feasibility of covert communications when cast as an evasion attack against a DL-based classifier employed by an eavesdropper. 

\bibliographystyle{IEEEtran}
\bibliography{lib}

\end{document}